\documentclass[aps,prl,showpacs,twocolumn,superscriptaddress, floatfix, tightenlines, amsmath, amssymb,longbibliography]{revtex4-1}

\usepackage{graphicx}
\usepackage{dcolumn}
\usepackage{bm}
\usepackage{xcolor} 
\usepackage{url}

\renewcommand{\bf}[1]{\textbf{#1}}

\begin{document}


\title{Nanoscale electronic inhomogeneities in  1\textit{T}-TaS$_2$}

\author{Campbell, B.}
\thanks{These authors contributed equally}
\affiliation{Department of Physics and Astronomy, University of New Hampshire, Durham, NH 03824.}
\author{Riffle, J. V.}
\thanks{These authors contributed equally}
\affiliation{Department of Physics and Astronomy, University of New Hampshire, Durham, NH 03824.}
\author{de la Torre, A.}%
\thanks{Current address: Department of Physics, Northeastern University, Boston, MA 02115}
\affiliation{Department of Physics, Brown University, Providence, RI, 02912.}
\author{Wang, Q.}%
\affiliation{Department of Physics, Brown University, Providence, RI, 02912.}
\author{Plumb, K. W.}%
\affiliation{Department of Physics, Brown University, Providence, RI, 02912.}
\author{Hollen, S. M.}%
\affiliation{Department of Physics and Astronomy, University of New Hampshire, Durham, NH 03824.}
\email{shawna.hollen@unh.edu}

\date{\today}

\begin{abstract}
We report a set of scanning tunneling microscopy (STM) and spectroscopy (STS) experiments studying native defects in CVT grown 1\textit{T}-TaS$_2$. Six different sample surfaces from four bulk crystals were investigated. Wide area imaging reveals a prevalence of nanometer-scale electronic inhomogeneities due to native defects, with pristine regions interspersed. These inhomogeneities appear in typical as-grown crystals and coexist with a well-formed commensurate charge density wave of 1\textit{T}-TaS$_2$ at low temperatures.
Electronic inhomogeneities show up both as variations in the apparent height in STM and in the local density of states in STS; the bands can shift by 60 meV and the gap varies by more than 100 meV across inhomogeneities.
These inhomogeneities are present in similar concentration across large-scale areas of all samples studied, but  do not influence the charge density wave formation on local or global scales. 
The commensurate charge density wave exhibits long-range order and remains locally intact in the presence of these inhomogeneities.

\end{abstract}

\maketitle

\section{Introduction}
Lattice defects can greatly affect the structural, optical, and electronic properties of materials and have an increased impact in the 2D limit, where much materials and device development is now focused. Of the dichalcogenides, 1\textit{T}-TaS$_2$ is an exciting material because of its rich phase diagram, complex examples of unusual phenomena, and potential applications in memory and ultra-fast switching devices \cite{Manzke:1989,Xu:2021,Mihailovic2021,Bauers2021,Huh2020,Yoshida2015}. The many reports describing observations of and theoretical explanations for unusual behavior in 1\textit{T}-TaS$_2$--including Mott insulation, quantum spin liquid behavior\cite{Law2017}, memristive switching\cite{Mihailovic2021,Yoshida2015}, hidden and metastable phases\cite{Stojchevska2014,Vaskivskyi2015}, and topological and chiral charge density waves\cite{Ravnik2019,Gao2021}--demonstrate its complexity and the level of interest from the condensed matter community. An important but often overlooked point in this discussion is the impact of native defects on the properties of 1\textit{T}-TaS$_2$. Induced defects were shown to suppress the commensurate CDW phase and insulating ground state and induce superconductivity at 2.5 K in one example \cite{Li2013}, and in another, K dopants did not affect the CDW order, but did induce metallicity\cite{Zhu2019}. On the other hand, intrinsic defects have gained limited attention. A very recent study reported the local electronic structure of intrinsic defects by STM and characterized 5 distinct defects by local density of states measurements and made partial assignment of these defects using density functional theory calculations \cite{Lutsyk2023}. 

We report a set of low-temperature scanning tunneling microscopy (STM) experiments over large areas of 1\textit{T}-TaS$_2$ that reveal the existence of nanometer-scale electronic inhomogeneities in addition to the commonly observed commensurate charge density wave (C-CDW).
We surveyed 6 different sample surfaces from 4 bulk crystals, and investigated large areas of each sample.
The inhomogeneities are observed as variations in the apparent height in STM topographs and as variations in the local density of states (LDOS) measured by scanning tunneling spectroscopy (STS). STS shows that the band edges shift by up to 60 mV across these features and the gap is suppressed near their center.  
Atomic resolution images support that lattice defects are a source of the electronic inhomogeneity. While the local CDW amplitude is affected by the electronic inhomogeneity, we find that the period and phase of the C-CDW are not modified by the defects or the associated electronic inhomogeneities. 
 Since the bulk features of these samples, including the resistivity versus temperature and the observation of the C-CDW by XRD, are in line with broadly accepted results from the literature, the coexistence of nanoscale inhomogeneities with the C-CDW demonstrates the importance of real space images of 1\textit{T}-TaS$_2$ and similar materials, and could potentially contribute to understanding their perplexing behavior. 

\begin{figure}[!b]
    \centering
    \includegraphics[width=\columnwidth]{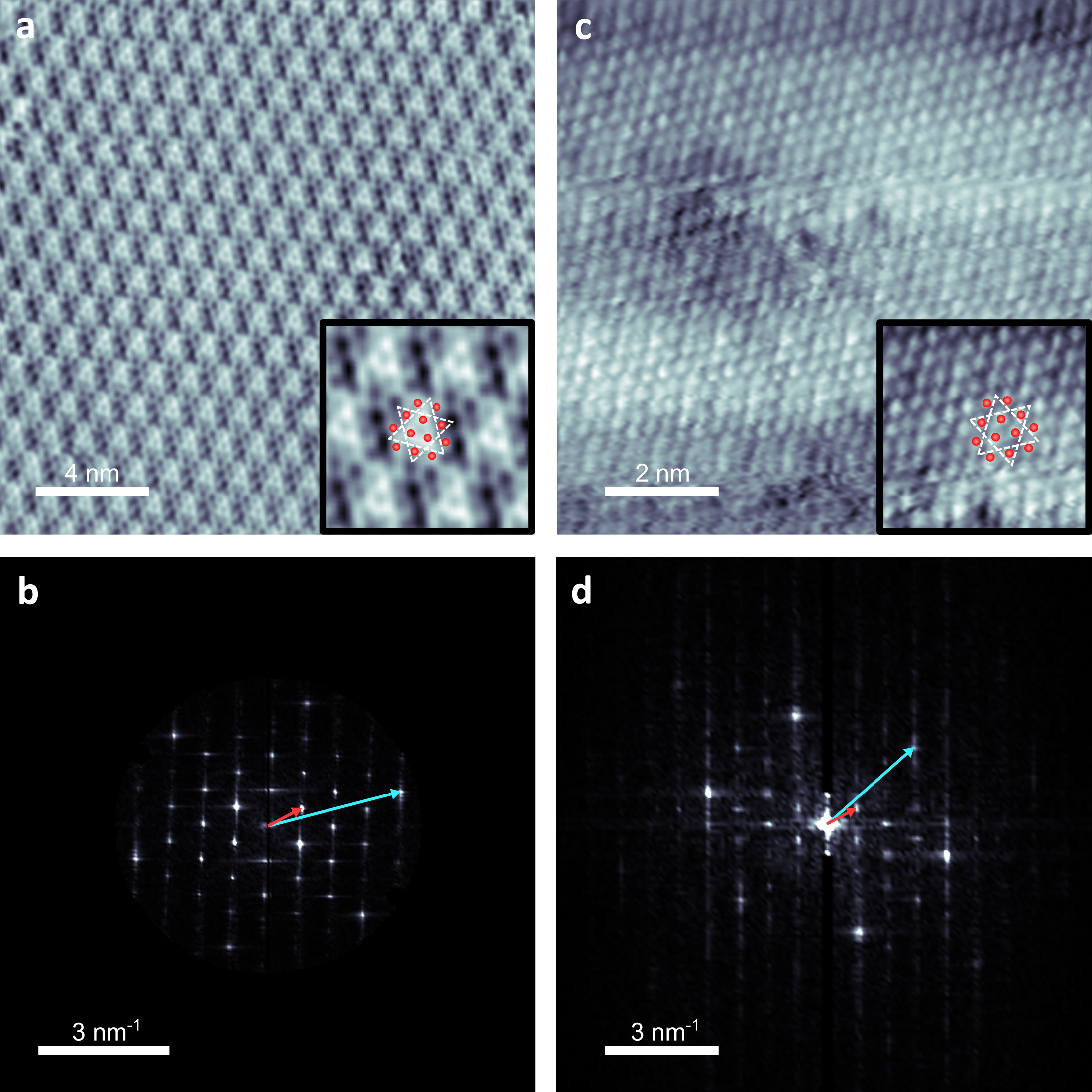} 
    \caption { \footnotesize
        STM topographs of the  1\textit{T}-TaS$_2$ surface at $T = 10$ K after cleaving in ultrahigh vacuum.
        \bf{a, c)}. Topography showing the C-CDW and atomic resolution ((a) 500 mV, 150 pA; (b) 550 mV, 145 pA). Insets: magnified view showing resolution of S atoms with lattice overlay (red dots) and star-of-David C-CDW pattern (dashed white triangles). \bf{b, d)} fast-Fourier transform (FFT) of images in {a}, {c}. Blue vector corresponds to $q_{\text{lattice}}$ and red vector to $q_{\text{CDW}}$. $\vert q_{\text{CDW}}\vert = 0.853 \pm 0.03 ~\text{nm}^{-1}$ ($0.27 \pm 0.12$~rlu) for (b), $\vert q_{\text{CDW}}\vert = 0.912 \pm 0.11 ~\text{nm}^{-1}$ ($0.31 \pm 0.25$~rlu) for (d).
    } \label{fig:control_comparison}
\end{figure}

\section{Experimental Details}
Single crystals of 1\textit{T}-TaS$_2$ were grown from stoichiometric amounts of elemental Ta and S by chemical vapor transport, using iodine as a transport agent.  Starting materials were sealed in quartz tubes and heated in a three zone furnace under a $950\textrm{C} - 850\textrm{C}$ temperature gradient for 240h, and then quenched in ice water to stabilize the \textit{1T} phase. 

To perform the STM experiments, a bulk TaS$_2$ crystal was mounted onto a stainless steel STM sample plate using conductive, UHV-safe epoxy, then introduced into the ultrahigh vacuum chamber. Using a cleaving screw and carbon tape, the surface of the TaS$_2$ crystal was cleaved at room temperature and ${\sim}10^{-10}$ torr.
The sample was then studied using our RHK Technology PanScan Freedom, closed-cycle STM, with an operating temperature of 10 K.
Pt-Ir cut tips were used in all the experiments reported here.
Images were analyzed with WSxM\cite{Horcas2007} and custom python codes available on github \cite{campbell}.
Determination of the band gap for each STS spectrum was done by identifying at least 6 contiguous dI/dV measurements that were within 0.25 pS of zero. 
The gap width was then defined as the voltage range of this contiguous set, and the gap center was defined as the average of the end points of this contiguous set.
The tolerance of 0.25 pS was fine-tuned by visual inspection of the resulting gaps plotted over their respective LDOS curve. 

\section{Experimental Results}
We surveyed six crystal surfaces across four different bulk crystals of  1\textit{T}-TaS$_2$, all grown by a standard chemical vapor transport method (see methods) and cleaved in ultrahigh vacuum at room temperature. Bulk resistivity measurements of as-grown crystals match expectations from literature and temperature dependent x-ray diffraction reveals well formed charge-density waves exhibiting a nearly-commensurate to commensurate transition at 190~K and significant thermal hysteresis consistent with literature.\cite{Fazekas1979,Fazekas1980} When surveying large area regions, we found significant populations of defects (approximate density of 300 per million atoms) and signatures of electronic disorder, with pristine regions interspersed, using STM and STS at 10~K. 
Figure~\ref{fig:control_comparison} presents images of two different samples from these experiments.
Figure~\ref{fig:control_comparison}a shows an STM topographic image of a pristine region of a 1\textit{T}-TaS$_2$ sample with uniform contrast while Fig.~\ref{fig:control_comparison}c shows a similar image from a region with nanometer-scale inhomogeneities. Figure~\ref{fig:control_comparison}a is more similar to typically reported STM images. It shows bright features in a triangular lattice with a periodicity of 1.2 nm arising from the commensurate charge density wave (C-CDW), and it shows the atomic lattice of the surface S atoms (see overlay in inset).
The C-CDW in STM topographs is primarily an electronic feature with an apparent height that corresponds to the integrated local density of states between the tip-sample bias and the Fermi energy. It is commensurate with the atomic lattice, forming a $\sqrt{13} \times \sqrt{13}$ superstructure rotated 13.9$^{\circ}$ from the  lattice vectors. \cite{Wilson1975,Fazekas1979, Fazekas1980}
In order to extract the C-CDW wavevector, we compute the fast-Fourier transform (FFT) of the topographic image (Fig.  ~\ref{fig:control_comparison}b). By comparing the magnitude and direction of the $q$ vector for the CDW (inner peaks, red arrow) and the atomic lattice (outer peaks, blue arrow), we extract a C-CDW wavevector that is consistent with the expected values for the C-CDW.

Figures~\ref{fig:control_comparison}c and \ref{fig:control_comparison}d show the same type of data as Fig.~\ref{fig:control_comparison} a and b, but for a region with inhomogeneities.
The C-CDW and atomic lattice are both resolved and the FFT compares well with \ref{fig:control_comparison}b, showing both the atomic lattice and the CDW.
There are differences between these images that can be attributed to the tip termination.
For example, the atomic lattice in Fig.~\ref{fig:control_comparison}c is clearly resolved, but the charge density wave is much more subtle than in \ref{fig:control_comparison}a (see star of David in Fig.~\ref{fig:control_comparison}c inset).
The background contrast in~\ref{fig:control_comparison}c is not uniform, and this nonuniformity is present for a wide range of imaging parameters.
There is a dark region left of center, roughly 5 nm across, and another in the upper right corner. These features  create a diffuse signal centered on zero in the FFT (~\ref{fig:control_comparison}d).
Supplemental Figure 1 shows a representative set of images that all show a clear C-CDW with an inhomogeneous background from 4 different surfaces from 3 different bulk crystals.
Figure~\ref{fig:control_comparison}a, is a region of Supplemental Fig. 1a that is free of defects.

\begin{figure}
    \centering
    \includegraphics[width=1\columnwidth]{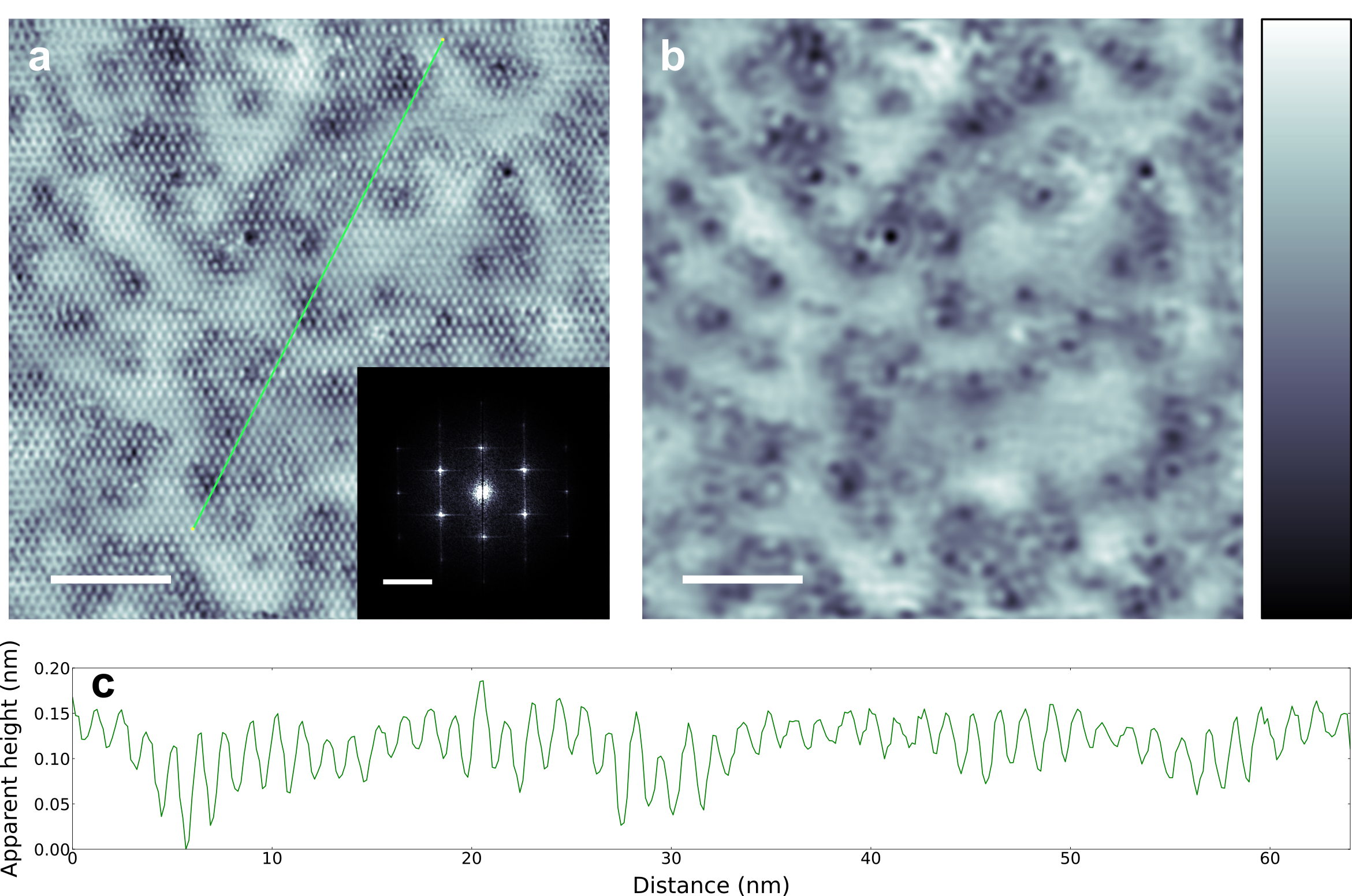} 
    \caption { \footnotesize
        \textbf{a)} Large area STM topograph showing the C-CDW and apparent height variations (300 mV, 65 pA setpoint). Scale bar is 14 nm.
        Inset: fast-Fourier transform. FFT scale bar is 1.5 nm$^{-1}$. 
        \textbf{b)} Topograph with C-CDW filtered out to just show the large-scale apparent height modulations. Color bar range is from 0 nm to 0.246 nm. 
        \textbf{c)} Line cut over green line in (a) showing the 1.2 nm periodic modulations of the C-CDW and the additional large-scale structure and amplitude variations created by the larger-scale electronic disorder.
    } \label{fig:large_w_line_cut}
\end{figure}

Large area images, like those in Fig. ~\ref{fig:large_w_line_cut}, show that inhomogeneities are distributed over the entire surface. These nanometer-scale inhomogeneites coexist with the C-CDW across the same spatial regions as clearly shown in the  $70 \times 70$ nm$^2$ image in Fig. ~\ref{fig:large_w_line_cut}. The FFT of this region exhibits sharp peaks with a sixfold symmetry, indicating the triangular lattice of the C-CDW is well-ordered and single-phased.
A linecut in Fig.~\ref{fig:large_w_line_cut}c reveals both the 1.2 nm periodicity of the C-CDW and the non-periodic modulations in amplitude that are due to the 5-10 nm-diameter inhomogeneities in the apparent height.
Fig. ~\ref{fig:large_w_line_cut}b shows the same area after applying a cut-off filter to remove the C-CDW peaks and emphasize the spatial distribution and variations among the inhomogeneities.
The inhomogeneities are centered around regions of low apparent height, which for these imaging parameters varies by 0.2 nm on average.

\begin{figure}[!t]
    \centering
    \includegraphics[width=\columnwidth]{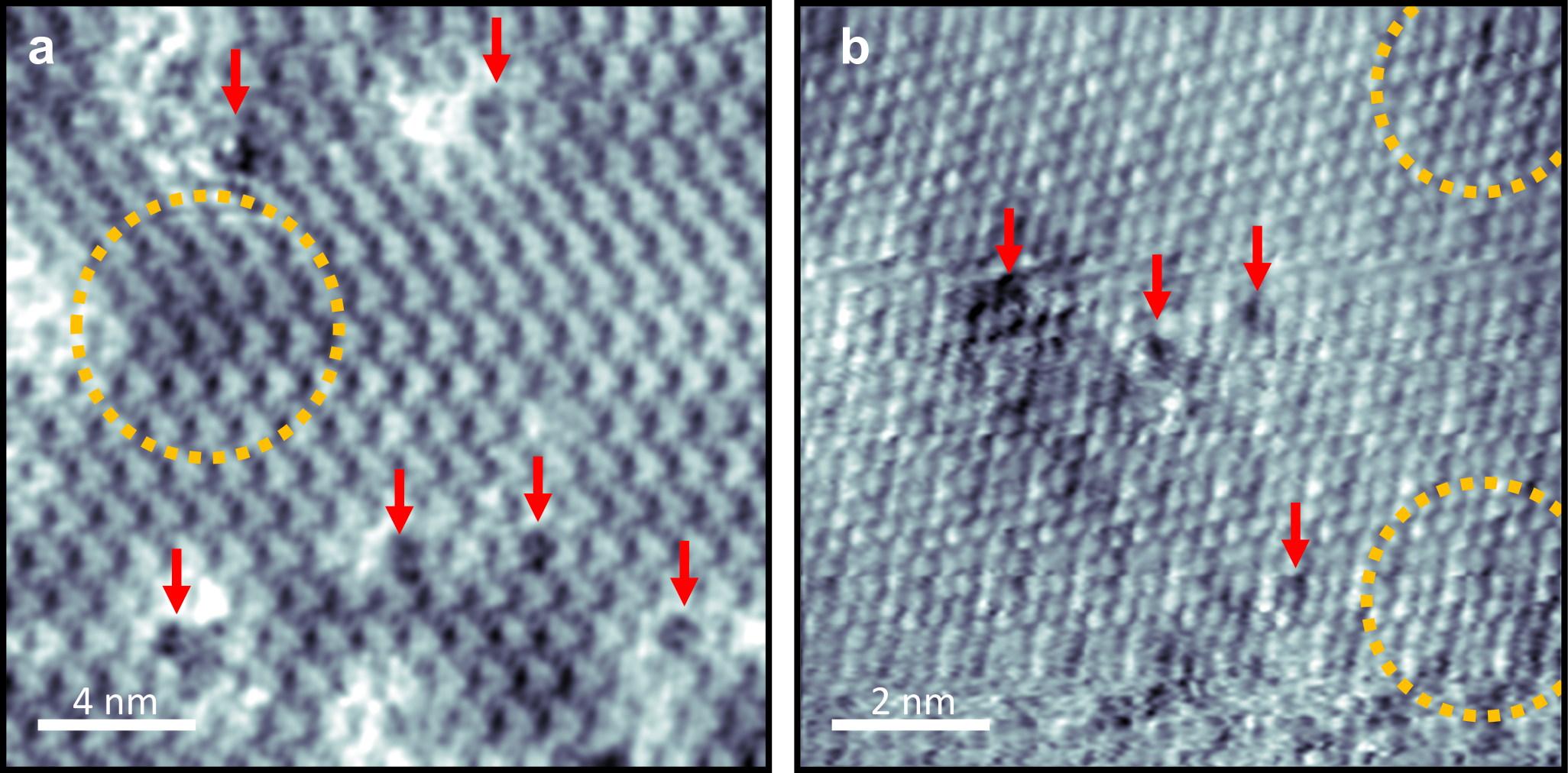} 
    \caption { \footnotesize
        STM topographs showing the C-CDW and defective atomic lattice.
       \textbf{a)} -500 mV and 50 pA and \textbf{b)} 550 mV and 145 pA.
        Red arrows indicated regions where the atomic lattice appears disrupted.
        Dashed yellow circles highlight regions with diffuse apparent height differences, but no lattice disruption.
    } \label{fig:atomic_w_red_arrows}
\end{figure}

To gain more insight into the source of these inhomogeneities, we examine atomically resolved images, like those in Figure~\ref{fig:atomic_w_red_arrows}. Here we see the atomic lattice appears to be disrupted near several bright and dark features, as indicated by red arrows in the figure.
Based on these and similar images, we assign the source of the inhomogeneities to lattice defects, most likely vacancies and substitutions.
There are also regions in Fig.~\ref{fig:atomic_w_red_arrows} that show contrast in the apparent height, but no obvious disruption to the surface atoms (dashed yellow circles).
These features appear larger by a factor of $\approx 1.5-2$ across and are less pronounced than those associated with surface defects.
Larger scale images (Fig.~\ref{fig:large_w_line_cut}) also show nanometer-scale features with a range of apparent heights and lateral sizes. These observations are all consistent with defects imaged in multiple crystalline layers below the surface, a common occurrence in STM imaging of semiconductors \cite{Feenstra1993,ebert1999,wong2015}.
Thus, defects that cause the nanoscale electronic inhomogeneities are not limited to the surface layer (so are not a product of the cleave), and are likely evenly distributed throughout the bulk crystal.

\begin{figure*}[!ht]
    \centering
    \includegraphics[width=\textwidth]{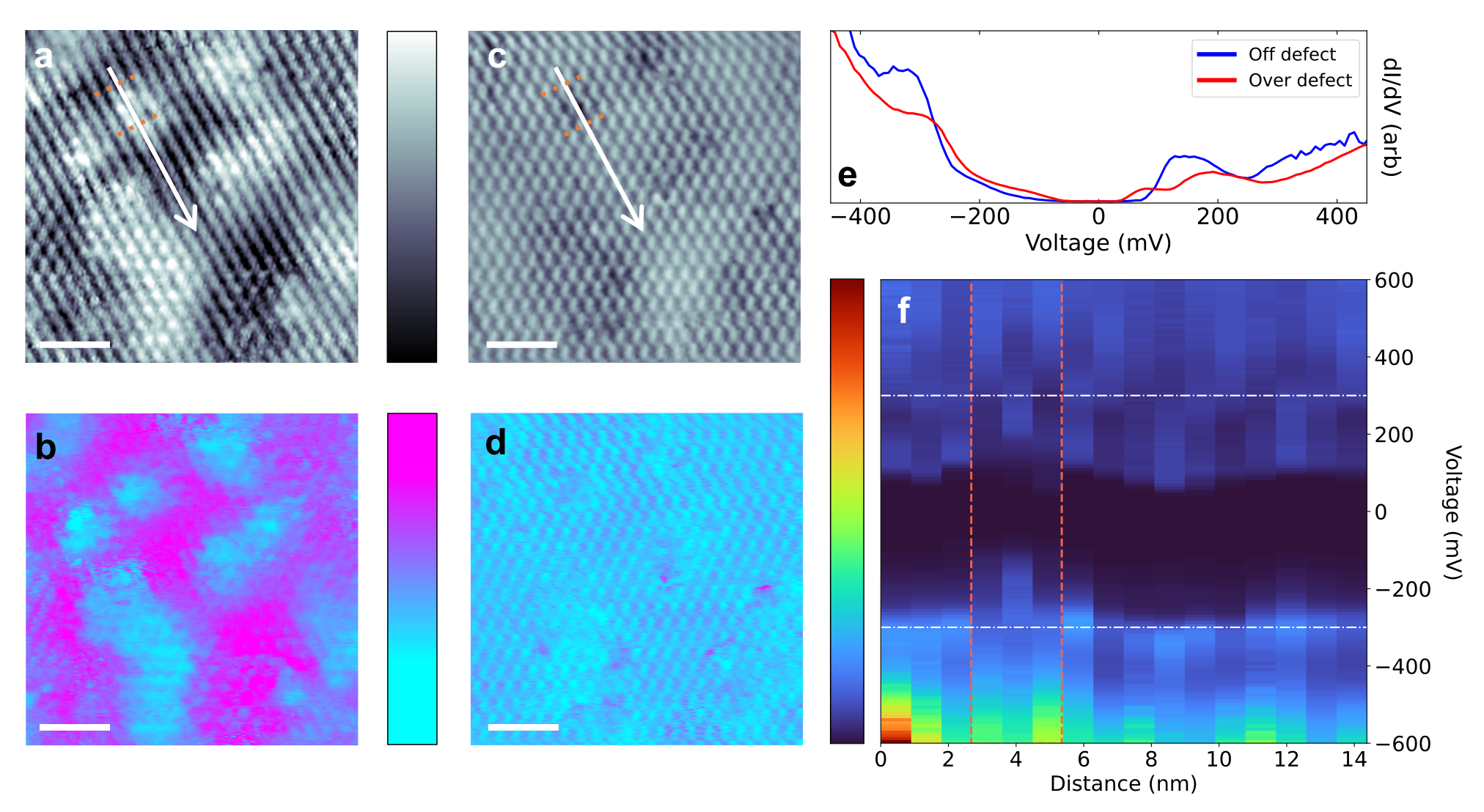} 
    \caption { \footnotesize
        \textbf{a, c)} STM topographs and \textbf{b, d)} dI/dV maps taken at -300 mV and 300 mV, respectively. Scale bar is 5 nm. 
\textbf{e)} dI/dV spectroscopy taken on and off the defect between the orange dashed lines in {a} and {c}.
        \textbf{f)} dI/dV point spectroscopy taken over the white line marked in a and c. Orange, dashed lines denote the boundary of the defect and the x-axis denotes distance along the cut, increasing in the direction of the arrow in {a, c}. The horizontal, white lines denote slices at $\pm 300$ mV that correspond to the dI/dV maps in b, d.
    } \label{fig:STS_line_plot}
\end{figure*}

The defects have a strong impact on the local electronic structure of the crystal, which we illuminate with measurements of the local density of states (LDOS).
First, by comparing topography at negative bias (filled states, Fig.~\ref{fig:STS_line_plot}a) and positive bias (empty states, Fig.~\ref{fig:STS_line_plot}c) we observe a strong voltage dependence of the apparent height, which indicates significant variation in the LDOS. 
We directly probed the LDOS using both dI/dV spatial mapping at set biases (Fig.~\ref{fig:STS_line_plot}b, d) and location-dependent spectroscopy (Fig.~\ref{fig:STS_line_plot}e, f). The spatial features of the LDOS in Fig.~\ref{fig:STS_line_plot}b, d mirror those in the topography, Fig.~\ref{fig:STS_line_plot}a, c.
At the chosen positive bias, the LDOS variations are weaker and more localized (Fig.~\ref{fig:STS_line_plot}d) than those at the chosen negative bias (Fig.~\ref{fig:STS_line_plot}b).

\begin{figure}
    \centering
    \includegraphics[width=1\columnwidth]{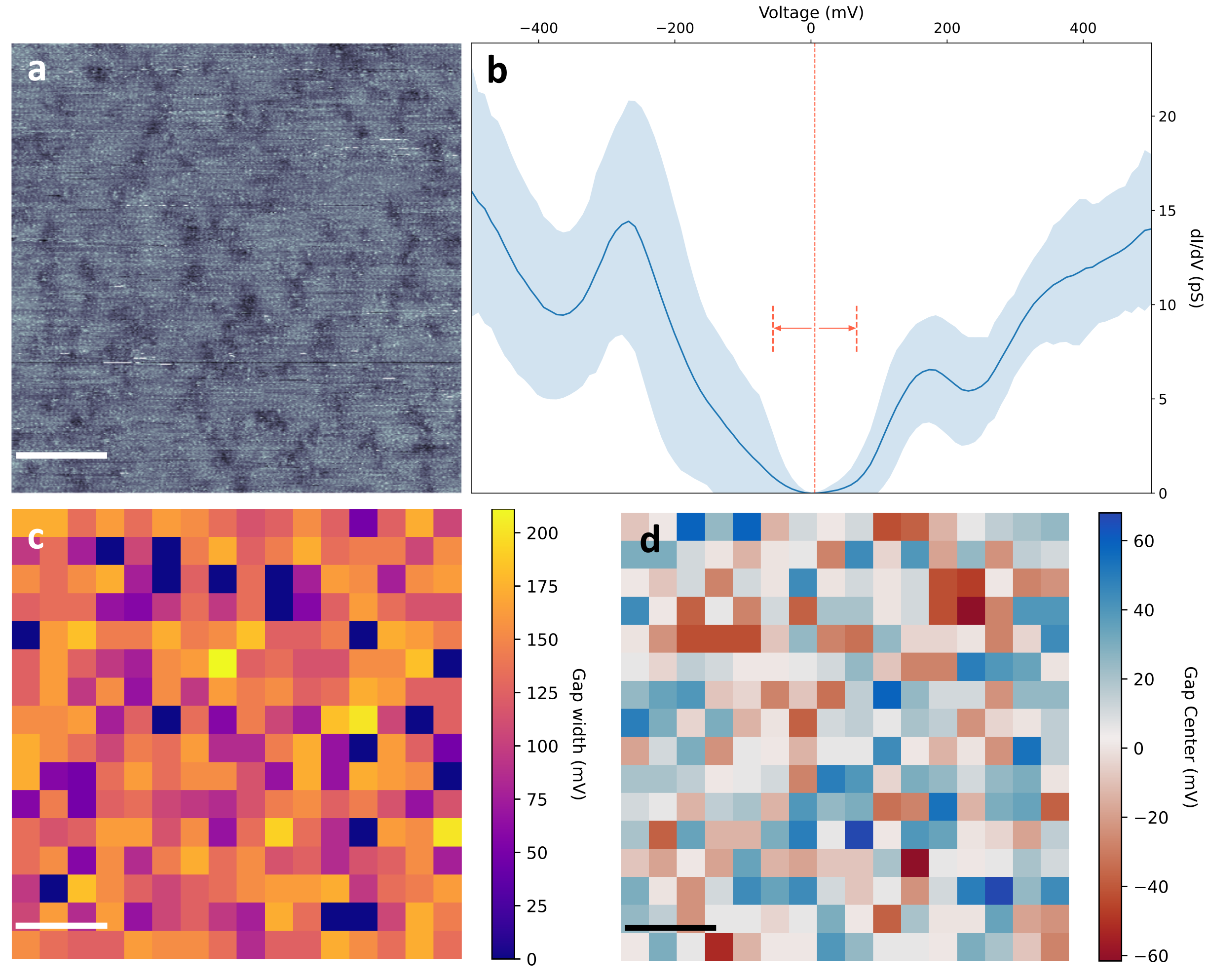} 
    \caption { \footnotesize 
        $16 \times 16$ grid of STS measurements over a 100 $\times$ 100 nm$^2$ area showing nanometer-scale inhomogeneities in gap width and gap center. 
        \bf{a)} STM topograph showing the nanometer-scale disorder.
        \bf{b)} Mean LDOS taken over all $16 \times 16$ spectra plotted in blue with standard deviation denoted by the shaded region. The mean gap center is marked by the red vertical, dashed line and the mean gap width is indicated by the red arrows. Mean gap width is $120 \pm 50$ mV and gap centers show variations up to 60mV.
        \bf{c)} Spatial plot of the gap width as defined in methods. Each pixel represents an STS measurement at the center of a $6.25 \times 6.25$ nm$^2$ section of \bf{(a)}.
        \bf{d)} Spatial plot of the gap center as defined in methods. Red and blue denote a negative and positive shift, respectively. 
        All scale bars are 20 nm.
    } \label{fig:gap_shifts}
\end{figure}

dI/dV point and line spectroscopy in Fig.~\ref{fig:STS_line_plot}e and f show the typical shape of the LDOS on 1\textit{T}-TaS$_2$ in which a $\sim$300 mV gap is bounded by two peaks, which are usually assigned to the upper and lower Hubbard bands of the Mott gap \cite{Kim1994,Cho2015}.
Here, we observe spatial shifts in the energy of the peaks and the gap edges of up to $\sim$ 60 mV. The relative contrast of the LDOS in Fig.~\ref{fig:STS_line_plot}b and d is also made clear since -300 mV (bottom dashed line in ~\ref{fig:STS_line_plot}f) cuts across the strong variations in the lower Hubbard band while +300 mV corresponds to an energy deeper in the conduction band, beyond the upper Hubbard band. 
Notably, the gap is suppressed over the defect in a and c (Fig. ~\ref{fig:STS_line_plot}e).  The combination of LDOS maps in Figs.~\ref{fig:STS_line_plot}b, d and the line spectra in Fig.~\ref{fig:STS_line_plot}e and f show that defects in the TaS$_2$ create strong electronic inhomogeneities, causing $\sim$ 60 mV variations in the local doping and strong disruptions to the local electronic structure. This observation is reminiscent of behavior in doped Mott insulators, including iridates \cite{Battisti2017} and cuprates \cite{PhysRevLett.93.097004, Cai2016}).

These strong local disruptions result in nanoscale inhomogeneities in the electronic structure over the entire crystal, as shown over a large region in Fig.~\ref{fig:gap_shifts}.
A $16 \times 16$ grid of dI/dV point spectroscopy indicate wide variations in both gap width and shifting of the gap center across a $100 \times 100 ~\textrm{nm}^2$ area.
The mean dI/dV spectra is shown in \ref{fig:gap_shifts}b as the solid blue LDOS curve with the standard deviation indicated by the shaded region.
The central red, vertical line shows the average gap center, and the red arrows indicate the average gap width. We chose to define the gap width using a threshold above $dI/dV=0$ (see Methods) so that it is independent of the upper and lower Hubbard peak locations. It should be noted that this choice does underestimate the gap width compared to the more typical choice that measures the separation of the Hubbard peaks. 
Fig.~\ref{fig:gap_shifts}c shows the spatial variations in gap width over the region in a; the average gap width is $120$ mV and the standard deviation of the set of gap widths is $50$ mV. 
The gap center in Fig.~\ref{fig:gap_shifts}d also shows large shifts, up to $60$ mV positive and negative. Over a large scale, the shifts tend to cancel, leading to an average shift close to zero (as indicated in Fig.~\ref{fig:gap_shifts}b).

\section{Discussion}

Because of the combined importance of physical and electronic structure, we cannot directly identify the defect types from STM topographs.
The defects imaged in Fig.~\ref{fig:atomic_w_red_arrows} seem to be mostly of the same species, but it is likely that over large areas we are imaging multiple types, including vacancies, substitutions, and possibly intercalants. Recent work by Lutsyk \textit{et al.}\cite{Lutsyk2023} identified 5 distinct defect types in high resolution STM an STS data, one of which they identified as a S vacancy, which are common in dichalcogenides\cite{Gao2021,Liang2021}. 
In atomic-resolution images, we are most sensitive to the outer S atoms of TaS$_2$, so this assignment appears to be consistent with data in Fig.~\ref{fig:atomic_w_red_arrows}. However, Lutsyk \textit{et al.} found the sulfur vacancy, identified by DFT, to have a very localized impact on the LDOS, extending no more than a single CDW site. Another defect, not identified in the DFT calculations but speculated to be a foreign atom substitution, was found to have electronic features extending several nanometers, which is more consistent with the large-area surveys we present here. Still this comparison is inconclusive since, in their study, the defect with nanometer-scale electronic imprint still exhibited a clear gap. 

Finally, we note that we do not see any signs of strain in our large scale images of these cleaved bulk crystals. Since strain is known to induce CDW domains and even a metallic mosaic phase \cite{Bu2019} and topological networks of CDW defects \cite{Ravnik2019}, lack of strain is consistent with our single-domain C-CDW observations. When combined with the LDOS measurements, this also makes clear that the nanoscale features are all electronic and do not correspond to bending or wrinkling of the surface. It is quite interesting that the C-CDW, which is known to be very sensitive to interlayer interactions(e.g. \cite{Lee2019}) as well as slight lattice strain, is robust to a high density of lattice defects and their resulting electronic disorder. This insensitivity of the CDW to lattice defects is observed in large scale STM images presented here, but also in bulk characterization of 1T-TaS$_2$: the C-CDW observed by XRD and the insulating character of the low temperature transport. This tension in the sensitivity of the C-CDW to interlayer interactions and slight strain but not defects deserves further study and could contribute to theory of the impact of disorder on charge density waves--a topic of significant recent interest\cite{Subires2023,Corasaniti2023,Xu2023,Baek2022}. These results also make clear that real space measurements are necessary to observe and study these structural and electronic inhomogeneities, which may play an important role in the observations of out-of-equilibrium states and potential applications like neuromorphic computing\cite{Yoshida2015}.

\section{Conclusions}
The data presented here show the prevalence and importance of nanoscale inhomogeneities to the local electronic properties of  1\textit{T}-TaS$_2$ at 10 K.
We show evidence that the inhomogeneities originate from lattice defects.
These features shift the band edges by up to 60 mV and in some cases the gap is degraded directly over the defect sites.
Overall these data contribute a broad picture of electronic inhomogeneities in  1\textit{T}-TaS$_2$ and CDW materials which will have an important impact on any electronic device applications and provides an example of a C-CDW robust to a high level of disorder.

\begin{acknowledgments}
Work performed at University of New Hampshire by B.C., J.V.R. and S.M.H. was supported by the National Science Foundation OIA 1921199, and a University of New Hampshire COVID recovery award. Work performed at Brown University by A.d.l.T., Q.W. and K.W.P. was supported by the U.S. Department of Energy, Office of Science, Office of Basic Energy Sciences, under Award Number DOE-SC0021265.
\end{acknowledgments}

\appendix

\bibliography{apsTaS2}

\end{document}